\documentclass[12pt]{article}
\usepackage{times}
\usepackage{geometry, parskip, hanging, hyperref, graphicx, floatrow}
\usepackage[export]{adjustbox}
\usepackage[labelfont=bf]{caption}
\usepackage{dblfnote}
\DFNalwaysdouble
\DeclareFloatVCode{afterfloat}{\vspace{-10pt}}
\usepackage[utf8]{inputenc}
\usepackage{enumitem,amssymb}
\usepackage{ragged2e}
\newlist{thematic}{itemize}{8}
\setlist[thematic]{label=$\square$}
\usepackage{pifont}
\newcommand{\cmark}{\ding{51}}%
\newcommand{\done}{\rlap{$\square$}{\raisebox{2pt}{\large\hspace{1pt}\cmark}}%
\hspace{-2.5pt}}

\newcommand{\Herschel}{\textit{Herschel}}
\newcommand{\Spitzer}{\textit{Spitzer}}

\newcommand{\SOFIA}{\textit{SOFIA}}

\begin{document}
{\raggedright
\huge
Astro2020 Science White Paper \linebreak

Time-Domain Photometry of Protostars at Far-Infrared and Submillimeter Wavelengths \linebreak
\normalsize

\noindent \textbf{Thematic Areas:} \hspace*{60pt} $\square$ Planetary Systems \hspace*{10pt} $\done$ Star and Planet Formation \hspace*{20pt}\linebreak
$\square$ Formation and Evolution of Compact Objects \hspace*{31pt} $\square$ Cosmology and Fundamental Physics \linebreak
  $\square$  Stars and Stellar Evolution \hspace*{1pt} $\square$ Resolved Stellar Populations and their Environments \hspace*{40pt} \linebreak
  $\square$    Galaxy Evolution   \hspace*{45pt} $\square$             Multi-Messenger Astronomy and Astrophysics \hspace*{65pt} \linebreak
  
\textbf{Principal Author:}

Name:	William J. Fischer
 \linebreak						
Institution:  Space Telescope Science Institute
 \linebreak
Email: wfischer@stsci.edu
 \linebreak
Phone:  667 218 6409
 \linebreak
 
\textbf{Co-authors:}

Michael Dunham (SUNY Fredonia),
Joel Green (STScI),
Jenny Hatchell (Univ.\ of Exeter),
Doug Johnstone (NRC Herzberg Astronomy and Astrophysics / Univ.\ of Victoria),
Cara Battersby (Univ.\ of Connecticut),
Pamela Klaassen (UK Astronomy Technology Centre),
Zhi-Yun Li (Univ.\ of Virginia),
Stella Offner (Univ.\ of Texas at Austin),
Klaus Pontoppidan (STScI),
Marta Sewi\l{}o (NASA Goddard),
Ian Stephens (Harvard-Smithsonian CfA),
John Tobin (NRAO),
Crystal Brogan (NRAO),
Robert Gutermuth (UMass Amherst),
Leslie Looney (Univ.\ of Illinois),
S.~Thomas Megeath (Univ.\ of Toledo),
Deborah Padgett (JPL),
Thomas Roellig (NASA Ames)
\linebreak

}

\textbf{Abstract:}

The majority of the ultimate main-sequence mass of a star is assembled in the protostellar phase, where a forming star is embedded in an infalling envelope and encircled by a protoplanetary disk. Studying mass accretion in protostars is thus a key to understanding how stars gain their mass and ultimately how their disks and planets form and evolve. At this early stage, the dense envelope reprocesses most of the luminosity generated by accretion to far-infrared and submillimeter wavelengths. Time-domain photometry at these wavelengths is needed to probe the  physics of accretion onto protostars, but variability studies have so far been limited, in large part because of the difficulty in accessing these wavelengths from the ground. We discuss the scientific progress that would be enabled with far-infrared and submillimeter programs to probe protostellar variability in the nearest kiloparsec.

\pagebreak

{\bf Photometric Variability of Protostars as a Probe of Mass Accretion}

Stars form when dense cores in clouds of molecular gas collapse under their own gravity (Shu et al.\ 1987). For most stars, the first 500,000 years of collapse (Evans et al.\ 2009) feature a central hydrostatically supported object  embedded in a dusty circumstellar envelope. Envelope material falls onto a protoplanetary disk, which supplies the matter needed for the star to reach its ultimate main-sequence mass. Some fraction of the disk material forms planets. We define the protostellar phase as that in which the envelope and disk still play an important role in the observed properties and subsequent evolution of the system. Studies of large samples of protostars reveal the conditions under which our own solar system formed. {\bf Here we discuss our current understanding of protostellar accretion via time-domain photometry beyond 70 \boldmath$\mu$m and address what is needed in the 2020s and beyond to make further progress.} We focus on the formation of low-mass stars ($\lesssim 1~M_\odot$); a contribution by Hunter et al.\ discusses variable accretion in massive protostars.

\begin{figure}
\includegraphics[trim={6 8 13 15},clip,width=\textwidth]{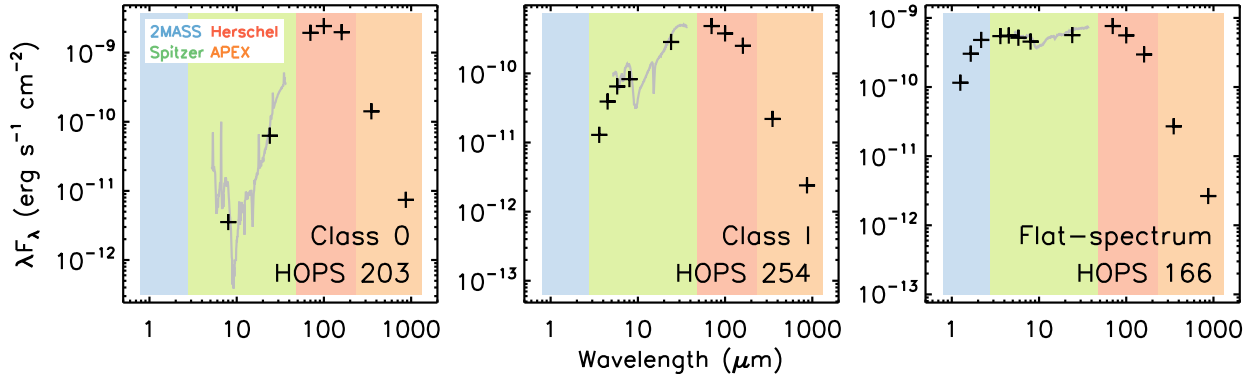}
\caption{SEDs of Orion protostars from Furlan et al.\ (2016) that demonstrate the full range of protostellar evolutionary classes, from least evolved (left) to most evolved (right). We include photometry and spectroscopy from 2MASS (Skrutskie et al.\ 2006), \Spitzer\ (Werner et al.\ 2004), \Herschel\ (Pilbratt et al.\ 2010), and the Atacama Pathfinder Experiment (APEX; Güsten et al.\ 2006).}\label{f.seds}
\end{figure}

A protostellar spectral energy distribution (SED; see Figure~\ref{f.seds}) is expected to evolve to shorter wavelengths as envelope dust clears out over time and there is less material available for reprocessing and reradiating accretion luminosity. From least evolved to most evolved, Class 0 protostars are faint in the mid-infrared (mid-IR) and emit most of their radiation at far-IR wavelengths (André et al.\ 1993), Class~I protostars have SEDs that rise with wavelength in the mid-IR (Lada 1987), and flat-spectrum protostars have spectral energy distributions that are flat in the mid-IR (Greene et al.\ 1994). Even for the flat-spectrum protostars, the wavelength region beyond 70 $\mu$m is a major contributor to the luminosity.

After the protostellar phase, in T Tauri stars, accretion rates are of order 10$^{-8}$ $M_\odot$ yr$^{-1}$ (Hartmann et al.\ 2016). These rates persist for at most a few million years, so $\dot{M}\Delta t$ during this phase is of order only a few percent of the mass of the star. {\bf The vast majority of the mass assembly process must therefore take place in the protostellar phase.} The time dependence of protostellar accretion is a fundamental measurement of how stars gain their mass, and our understanding in this area is limited. Understanding this process will also enable insight into the origin of the initial mass function of stars. Furthermore, since the accretion process is determined by the physical and chemical properties within the disk, variability is intimately linked to the time-dependent structure and chemistry of the disk. Finally, accretion variability may lead to much of the observed structure within jets and outflows. {\bf In spite of dramatic progress made since the last decadal survey in the theory and observation of the earliest phases of star formation, our understanding of the physics of protostellar accretion remains poor.}

Discovering how accretion works in the youngest protostars will aid our understanding of how exoplanets form. Observations with submillimeter- to centimeter-wave interferometers have revealed that protostars frequently have disks (Tobin et al.\ 2012; Murillo et al.\ 2013; Yen et al.\ 2017; Segura-Cox et al.\ 2016, 2018), and studies of more evolved disks routinely reveal structure, suggesting that planets may have already formed in the protostellar phase (ALMA Partnership 2015; Long et al.\ 2018; Andrews et al.\ 2018). It remains to be seen whether the variability mechanisms that prevail in more evolved disks (Herbst et al.\ 1994) also operate in the disks of protostars.

\begin{figure}
\floatbox[{\capbeside\thisfloatsetup{capbesideposition={right,top},capbesidewidth=0.585\textwidth}}]{figure}[0.39\textwidth]
{\includegraphics[trim={146 0 180 0},clip,width=0.39\textwidth]{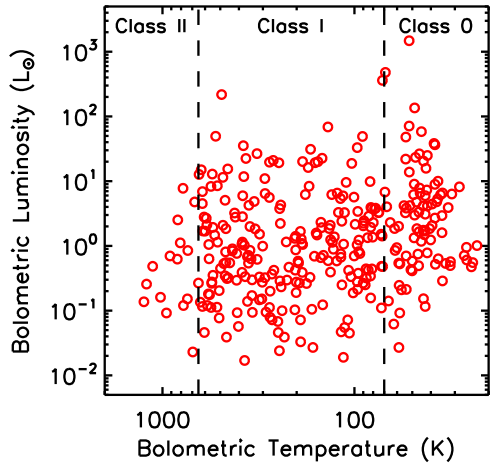}}
{\caption{Bolometric luminosities ($L_{\rm bol}$) and temperatures ($T_{\rm bol}$) of 330 young stellar objects in Orion, 315 of which have bolometric temperatures characteristic of protostars (Classes 0 and I). The flat-spectrum class is not included in this scheme, but it corresponds roughly to a range of 350--950 K (Evans et al.\ 2009). There is a broad spread in $L_{\rm bol}$ at each $T_{\rm bol}$, and the median luminosity is lower for larger $T_{\rm bol}$. {\bf The distribution is consistent with accretion luminosities that decline exponentially with time but are punctuated by episodes of enhanced accretion (Dunham et al.\ 2010; Fischer et al.\ 2017).}}\label{f.blt}}
\end{figure}

SEDs combining \Spitzer, \Herschel, and submillimeter data enabled the construction of bolometric luminosity and temperature (BLT) diagrams for hundreds of nearby protostars. This diagram was introduced by Myers \& Ladd (1993) as a protostellar equivalent to the Hertzsprung-Russell diagram, with the hope that it would lead to advances in the understanding of protostellar accretion similar to what was facilitated by HR diagrams for stellar physics. Unfortunately, BLT diagrams feature a spread of three orders of magnitude in luminosity at each temperature, without the clean main sequence that exists for middle-aged stars (Figure~\ref{f.blt}). Studies have investigated the role accretion plays in the broad trends seen in BLT diagrams (Dunham et al.\ 2010; Fischer et al.\ 2017), but the role of variable accretion in giving rise to the luminosity spread is not well quantified.

Class 0 and early Class I protostars, where the majority of the stellar mass is being assembled, are typically faint to invisible at near-IR wavelengths and shorter. Mid-IR surveys, however, routinely detect protostars and monitor their variability. The \Spitzer\ YSOVAR program (Morales-Calder\'on et al.\ 2011) obtained 3.6 and 4.5 $\mu$m light curves for protostars in various star-forming regions. The origins of protostellar variability are complex at these wavelengths; variable disk structures as well as accretion play a role (Rebull et al.\ 2015; Wolk et al.\ 2018). Protostars are brightest in the far IR, which probes heating of the inner envelope by accretion processes. This wavelength range is thus the ideal one for tracking accretion-driven variations in the luminosity. Beyond the far-IR, changes in the submillimeter are more sensitive to the temperature in the outer envelope, where interstellar heating is a factor (Johnstone et al.\ 2013). Thus, while the far-IR and submillimeter are the critical wavelengths for this work, near- and mid-IR observations have important supporting~roles.

{\bf Evidence for Variability in Protostars beyond 70 \boldmath$\mu$m}

Young stars may undergo episodic accretion outbursts, where the luminosity may increase by a factor of more than 100 and remain high for decades (Hartmann \& Kenyon 1996). In recent decades, protostellar outbursts have been discovered (Strom \& Strom 1993; Caratti o Garatti et al.\ 2011; Safron et al.\ 2015), and fossil evidence has been found for previous outbursts in the form of extended C$^{18}$O emission due to past heating (J\o rgensen et al.\ 2015; Frimann et al.\ 2017). Outbursts are thought to be due to the interplay between gravitational and magneto-rotational instabilities in the disk, which are mediated by infall from the envelope (e.g., Zhu et al.\ 2012; Vorobyov \& Basu 2015; Kratter \& Lodato 2016). However, the origin of accretion variations and the nature of angular momentum transport are debated. The outburst mechanism will be constrained by future studies focusing on the physical and chemical properties of disks (see the contribution by Sheehan et al.).

\begin{figure}
\includegraphics[trim={57 0 110 0},clip,width=0.6\textwidth]{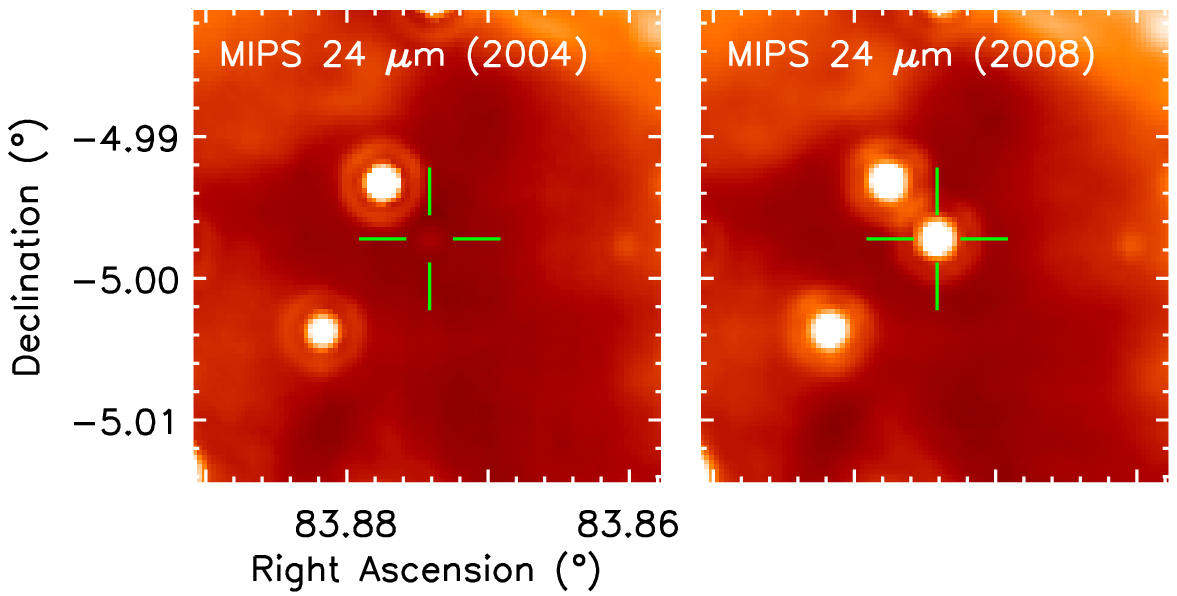}\includegraphics[trim={74 0 135 0},clip,width=0.4\textwidth]{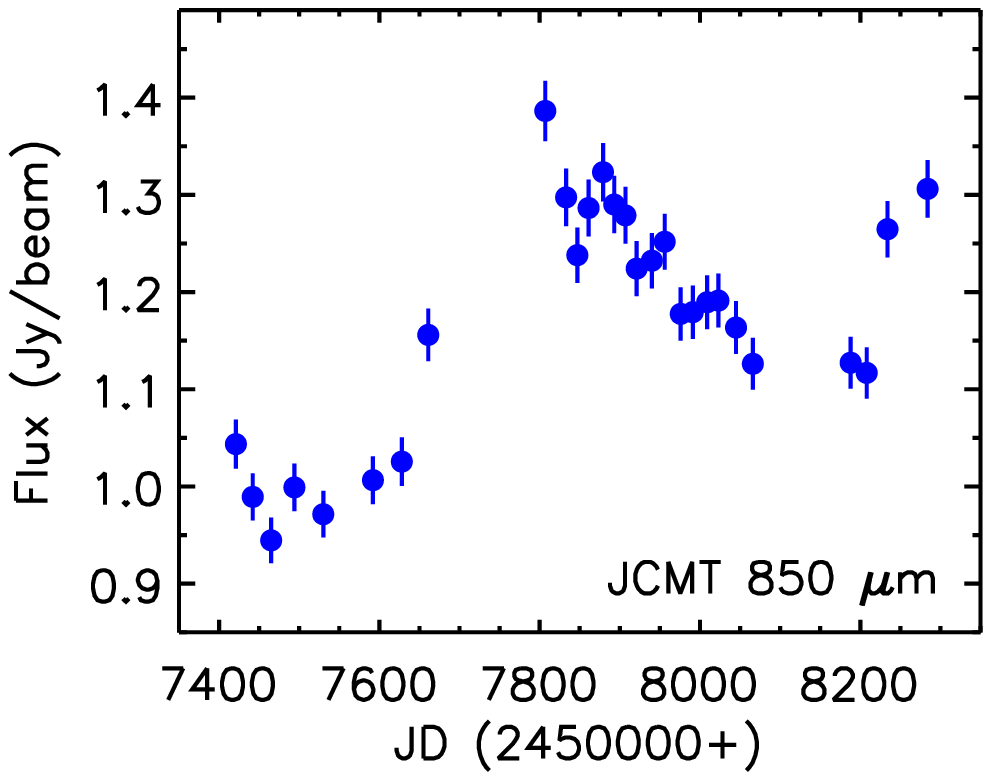}
\caption{{\em Left:} \Spitzer\ 24 $\mu$m images that demonstrate the outburst of the Class 0 protostar HOPS~383 in Orion (Safron et al.\ 2015). {\em Right:} JCMT 850 $\mu$m light curve of the quasi-periodic Class I protostar EC~53 in Serpens Main (Yoo et al.\ 2017; Johnstone et al.\ 2018b).}\label{f.h383}
\end{figure}

In 319 Orion protostars, Fischer et al.\ (2019) reported two outbursts over 6.5 yr, with increases in luminosity by factors of $\sim$ 12 and 40. This result and the incidence of bursting sources elsewhere (Offner \& McKee 2011) point to a 1000~yr interval between bursts of a given protostar, but with large uncertainty, comparable to estimates from jet structure and proper motions (Raga et al.\ 1990). The left panels of Figure~\ref{f.h383} demonstrate the outburst of HOPS 383, the first known Class 0 outburst (Safron et al.\ 2015). Far-IR surveys will uncover additional deeply embedded outbursts, reduce the uncertainty on their occurrence rate (Hillenbrand \& Findeisen 2015), and perhaps detect the signature of the beginning of accretion in the youngest protostars.

Smaller-scale far-IR variations have also been detected in protostars. Billot et al.\ (2012) obtained \Herschel/PACS photometry of 43 protostars in a 35$^\prime$ by 35$^\prime$ field in Orion at 70 and 160 $\mu$m. They found that protostars frequently varied by more than 20\% over periods as short as a few weeks, implying that accretion rates vary at least at this level. In the immediate future, \SOFIA\ has some capacity to track the variability of nearby, bright protostars at mid-to-far-IR wavelengths.

The JCMT Transient Survey is conducting similar monitoring across eight nearby star-forming regions at submillimeter wavelengths (Herczeg et al.\ 2017). It monitors more than 50 bright protostars with a monthly cadence and has discovered robust evidence of variability at 850 $\mu$m. Yoo et al.\ (2017) reported on a submillimeter burst in a protostar in Serpens Main (right panel of Figure~\ref{f.h383}). The burst phasing matches the protostar's 2 $\mu$m quasi-periodic light curve (Hodapp et al.\ 2012), suggesting a triggering mechanism within the inner few AU if the periodicity is related to orbital times.  Johnstone et al.\ (2018a) analyzed the first 18 months of survey data, finding that about 10\% of the surveyed protostars varied by about 5\% at 850 $\mu$m over the course of a year. 

{\bf Understanding Stellar Mass Assembly with Space-Based Far-IR Monitoring}

Far-IR monitoring of a large number of protostars will reshape our understanding of the physics of how stars get their mass and how this affects the formation of planets. While about 100 protostars have been monitored so far beyond 70 $\mu$m, there are about 1000 in the nearest kiloparsec, about 25\% of which are of Class 0 (Gutermuth et al.\ 2011; Dunham et al.\ 2015; Furlan et al.\ 2016). The monitoring could extend to extragalactic targets, especially for massive young stellar objects. In the \Spitzer\ SAGE survey of the Large Magellanic Cloud (Meixner et al.\ 2006), 1\% of massive protoclusters varied over three months (Vijh et al.\ 2009); the variability was likely driven by individual members of the clusters. Below we investigate the requirements to monitor those inside~1~kpc.

{\it Angular resolution:} An angular resolution of 5$^{\prime\prime}$ (obtainable at 100 $\mu$m with a 4--5 m telescope) corresponds to 5000 AU at 1 kpc. This is not fine enough to separate the components of typical multiple systems (Kounkel et al.\ 2016; Tobin et al.\ 2016), but it is sufficient to separate the systems from each other. In an unresolved system where one protostar shows dramatic variability, the fractional change in flux from the system will be reduced compared to that of the single protostar, but the detection can be investigated further with higher-resolution facilities.

\begin{figure}
\vskip 0.2in
\includegraphics[trim={94 0 115 0},clip,width=0.5\textwidth]{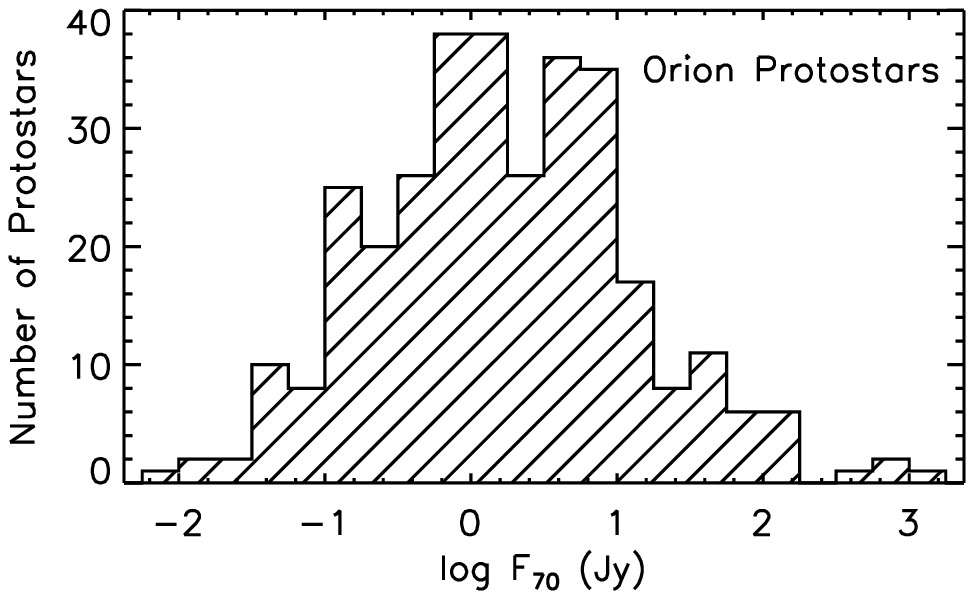}\includegraphics[trim={84 0 125 0},clip,width=0.5\textwidth]{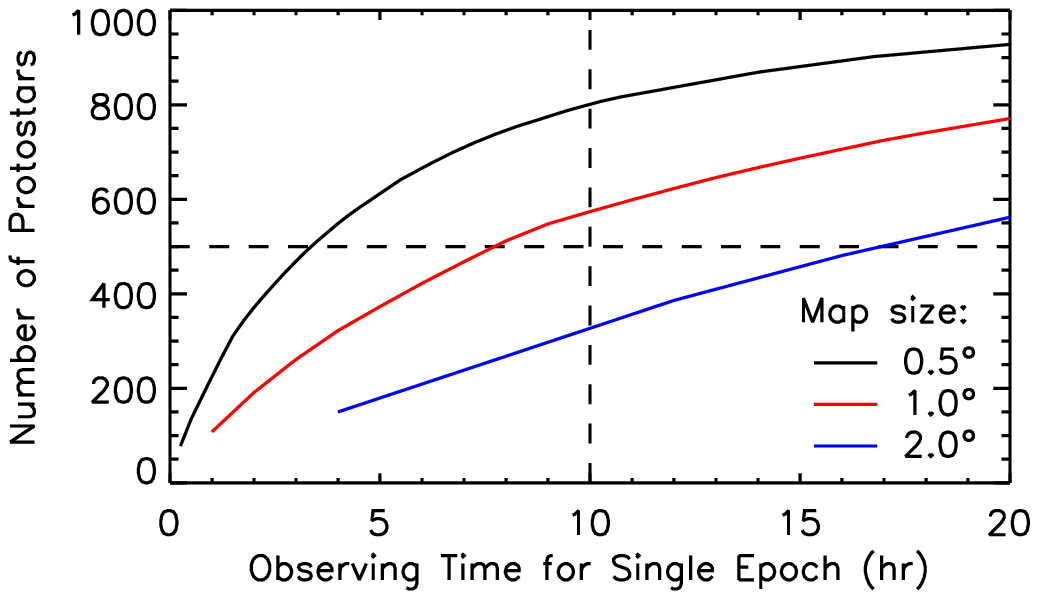}
\caption{{\em Left:} Histogram of 70 $\mu$m fluxes for 319 Orion protostars (Furlan et al.\ 2016). {\em Right:} Number of protostars in the nearest kpc that can be imaged at 1 deg$^2$ hr$^{-1}$ as a function of observing time and map size. The efficiency increases for smaller maps because protostars are clustered.}\label{f.f70}
\end{figure}

{\it Sensitivity:} As measured in \Herschel\ images, 75\% of the Orion protostars have 70 $\mu$m flux densities greater than 0.35 Jy (left panel of Figure~\ref{f.f70}). For protostars at 1 kpc, this corresponds to 60~mJy. We want to measure 5\% changes at three sigma, corresponding to a one-sigma sensitivity of $\sim$ 1~mJy to monitor such variations in nearly all protostars in the nearest kiloparsec. Such a survey would also be sensitive to submillimeter galaxies; we would distinguish between these and protostars with near-IR spectroscopy or imaging.

{\it Wavelength range:} Class I protostars have SEDs that peak near 70 $\mu$m (Figure~\ref{f.seds}). In \Herschel\ images, Stutz et al.\ (2013) discovered the reddest, most deeply embedded protostars in Orion.  (Two of these are submillimeter variables; see Johnstone et al.\ 2018a and Mairs et al.\ 2018.) These are so faint at 24 $\mu$m that they were not classified as protostars based on \Spitzer\ data alone. Their SEDs peak between 100 and 160 $\mu$m. Therefore, photometry between 70 and 160 $\mu$m is needed to recover the youngest protostars. Broad SEDs encompassing this range would be essential for disentangling the causes of variability in all regions of the envelope, disk, and central object.

{\it Mapping speed:} Mapping speed, not sensitivity, is the limiting factor for the science discussed here. \Herschel\ Gould Belt Survey archival data (Andr\'e et al.\ 2010) indicate that \Herschel's fastest mapping mode mapped about 1 deg$^2$ hr$^{-1}$. Based on the distribution on the sky of the protostars in the nearest kiloparsec, the right panel of Figure~\ref{f.f70} shows how the number of protostars that can be monitored at this speed depends on observing time and map size. Because protostars are clustered, smaller fields can be monitored more efficiently. Monitoring nearly all of these protostars would take $\lesssim$ 20 hr and could be repeated several times over the life of a mission. 

The parameters presented here are roughly consistent with those offered by the far-IR imager being considered for NASA’s proposed {\it Origins Space Telescope}\footnote{\url{https://origins.ipac.caltech.edu/}}. At the mapping speed considered here, {\it Origins} would also be able to obtain 25--590 $\mu$m spectral cubes at a resolving power of a few hundred. With these, one could compare the radiative energy balance of transients before and after their outbursts. The joint Japanese/European {\it SPICA}\footnote{\url{http://www.spica-mission.org/}} mission is expected to have similar capabilities, although with coarser angular resolution.

{\bf Understanding Accretion Physics with Ground-Based Submillimeter Monitoring}

Like with large-amplitude bursts, understanding the smaller-scale photometric variability of protostars has suffered from small sample sizes and short time baselines. The work of Billot et al.\ (2012) and the JCMT Transient Survey team have provided tantalizing evidence of far-IR and submillimeter variability, but the frequency and amplitude of this variability is not yet well constrained. CCAT-p\footnote{\url{http://www.ccatobservatory.org/}}, a planned 5 m submillimeter telescope dedicated to surveys, and AtLAST\footnote{\url{http://atlast-telescope.org}}, a proposed 50 m single-dish companion to ALMA being investigated by an international collaboration, would effectively map the requisite fields in the submillimeter, but for a longer period of time and more frequently than a space-based far-IR mission.

Submillimeter data are an essential part of multiwavelength monitoring during a burst, which tracks the progress of the temperature rise from the inside out, through the disk and envelope. This rise is reflected in the dust emission, affecting short wavelengths first and propagating to longer wavelengths as the extent of heating increases. The differences in lag time between wavelengths also reflect the physical geometry of the system. Structures close to the protostar are heated earlier and to higher temperatures. This heating is not instantaneous, as the time taken for photons to propagate outwards in such high optical-depth environments can be weeks or months; such timescales can be measured by comparing lightcurves at multiple wavelengths (Johnstone et al.\ 2013).

Time-domain photometry of large samples of protostars at wavelengths beyond 70 $\mu$m will enable major progress in understanding how stars gain their mass, how disks form and evolve, and how planets begin to emerge in the first few hundred thousand years of solar system formation. As the nature of photometric variability at these wavelengths becomes clear, followup spectroscopic studies will give additional insight into the responsible physics.

\pagebreak
\begin{hangparas}{0.25in}{1}
\raggedright
\textbf{References}\\
ALMA Partnership, Brogan, C. L., P\'erez, L. M., et al.\ 2015, ApJL, 808, L3\\
Andr\'e, P., Men'shchikov, A., Bontemps, S., et al.\ 2010, A\&A, 518, L102\\
Andr\'e, P., Ward-Thompson, D., \& Barsony, M. 1993, ApJ, 406, 122\\
Andrews, S. M., Huang, J., P\'erez, L. M., et al.\ 2018, ApJL, 869, L41\\
Billot, N., Morales-Calder\'on, M., Stauffer, J. R., Megeath, S. T., \& Whitney, B. 2012, ApJL, 753, L35\\
Caratti o Garatti, A., Garcia Lopez, R., Scholz, A., et al.\ 2011, A\&A, 526, L1\\
Dunham, M. M., Allen, L. E., Evans, N. J., II, et al.\ 2015, ApJS, 220, 11\\
Dunham, M. M., Evans, N. J., II, Terebey, S., Dullemond, C. P., \& Young, C. H. 2010, ApJ, 710, 470\\
Evans, N. J., II, Dunham, M. M., Jørgensen, J. K., et al.\ 2009, ApJS, 181, 321\\
Fischer, W. J., Megeath, S. T., Furlan, E., et al.\ 2017, ApJ, 840, 69\\
Fischer, W. J., Safron, E., \& Megeath, S. T. 2019, ApJ, 872, 183\\
Frimann, S., J\o rgensen, J. K., Dunham, M. M., et al.\ 2017, A\&A, 602, A120\\
Furlan, E., Fischer, W. J., Ali, B., et al.\ 2016, ApJS, 224, 5\\
Greene, T. P., Wilking, B. A., Andr\'e, P., Young, E. T., \& Lada, C. J. 1994, ApJ, 434, 614\\
G\"usten, R., Nyman, L. A., Schilke, P., et al.\ 2006, A\&A, 454, L13\\
Gutermuth, R. A., Pipher, J. L., Megeath, S. T., et al.\ 2011, ApJ, 739, 84\\
Hartmann, L., Herczeg, G., \& Calvet, N. 2016, ARA\&A, 54, 135\\
Hartmann, L., \& Kenyon, S. J. 1996, ARA\&A, 34, 207\\
Herbst, W., Herbst, D. K., Grossman, E. J., \& Weinstein, D. 1994, AJ, 108, 1906\\
Herczeg, G. J., Johnstone, D., Mairs, S., et al.\ 2017, ApJ, 849, 43\\
Hillenbrand, L. A., \& Findeisen, K. P.  2015, ApJ, 808, 68\\
Hodapp, K. W., Chini, R., Watermann, R., \& Lemke, R. 2012, ApJ, 744, 56\\
Johnstone, D., Hendricks, B., Herczeg, G. J., \& Bruderer, S. 2013, ApJ, 765, 133\\
Johnstone, D., Herczeg, G. J., Mairs, S., et al.\ 2018a, ApJ, 854, 31\\
Johnstone, D., Mairs, S., Naylor, T., et al.\ 2018b, ATel, 11614\\
J\o rgensen, J. K., Visser, R., Williams, J. P., \& Bergin, E. A. 2015, A\&A, 579, A23\\
Kounkel, M., Megeath, S. T., Poteet, C. A., Fischer, W. J., \& Hartmann, L. 2016, ApJ, 821, 52\\
Kratter, K., \& Lodato, G. 2016, ARA\&A, 54, 271\\
Lada, C. J. 1987, in Proc.\ IAU Symp.\ 115, Star Forming Regions, ed. M.\ Piembert \& J. Jugaku (Dordrecht:\ Reidel), 1\\
Long, F., Pinilla, P., Herczeg, G. J., et al.\ 2018, ApJ, 869, 17\\
Mairs, S., Bell, G. S., Johnstone, D., et al.\ 2018, ATel, 11583\\
Meixner, M., Gordon, K. D., Indebetouw, R., et al.\ 2006, AJ, 132, 2268\\
Morales-Calder\'on, M., Stauffer, J. R., Hillenbrand, L. A., et al.\ 2011, ApJ, 733, 50\\
Murillo, N. M., Lai, S.-P., Bruderer, S., Harsono, D., \& van Dishoeck, E. F. 2013, A\&A, 560, A103\\
Myers, P. C., \& Ladd, E. F. 1993, ApJL, 413, L47\\
Offner, S. S. R., \& McKee, C. F. 2011, ApJ, 736, 53\\
Pilbratt, G. L., Riedinger, J. R., Passvogel, T., et al.\ 2010, A\&A, 518, L1\\
Raga, A. C., Canto, J., Binette, L., \& Calvet, N. 1990, ApJ, 364, 601\\
Rebull, L. M., Stauffer, J. R., Cody, A. M., et al.\ 2015, AJ, 150, 175\\
Safron, E. J., Fischer, W. J., Megeath, S. T., et al.\ 2015, ApJL, 800, L5\\
Segura-Cox, D. M., Harris, R. J., Tobin, J. J., et al.\ 2016, ApJL, 817, L14\\
Segura-Cox, D. M., Looney, L. W., Tobin, J. J., et al.\ 2018, ApJ, 866, 161\\
Shu, F. H., Adams, F. C., \& Lizano, S. 1987, ARA\&A, 25, 23\\
Skrutskie, M. F., Cutri, R. M., Stiening, R., et al.\ 2006, AJ, 131, 1163\\
Strom, K. M., \& Strom, S. E. 1993, ApJL, 412, L63\\
Stutz, A. M., Tobin, J. J., Stanke, T., et al.\ 2013, ApJ, 767, 36\\
Tobin, J. J., Hartmann, L., Chiang, H.-F., et al.\ 2012, Nature, 492, 83\\
Tobin, J. J., Looney, L. W., Li, Z.-Y., et al.\ 2016, ApJ, 818, 73\\
Vijh, U. P., Meixner, M., Babler, B., et al.\ 2009, AJ, 137, 3139\\
Vorobyov, E. I., \& Basu, S. 2015, ApJ, 805, 115\\
Werner, M. W., Roellig, T. L., Low, F. J., et al.\ 2004, ApJS, 154, 1\\
Wolk, S. J., G\"unther, H. M., Poppenhager, K., et al.\ 2018, AJ, 155, 99\\
Yen, H.-W., Koch, P. M., Takakuwa, S., et al.\ 2017, ApJ, 834, 178\\
Yoo, H., Lee, J.-E., Mairs, S., et al.\ 2017, ApJ, 849, 69\\
Zhu, Z., Hartmann, L., Nelson, R. P., \& Gammie, C. F. 2012, ApJ, 746, 110\\
\end{hangparas}
\end{document}